\documentclass{aa} 
\usepackage{graphicx,txfonts}  

\begin{document}

\title{Simultaneous X-ray and Ultraviolet spectroscopy of the Seyfert
galaxy NGC~5548. III. X-ray time variability}

\authorrunning{J.S. Kaastra et al.} 
\titlerunning{X-ray time variability of NGC~5548}

\author{J.S. Kaastra\inst{1}
\and
K.C. Steenbrugge\inst{1}
\and 
D. M. Crenshaw\inst{2}
\and
 S. B. Kraemer\inst{3,4}
 \and
N. Arav\inst{5}
\and
 I. M. George\inst{6,7}
 \and 
D. A. Liedahl\inst{8}
\and
 R. L. J. van der Meer\inst{1}
 \and 
F. B. S. Paerels\inst{9}
\and
 T. J. Turner\inst{6,7}
 \and
  and T. Yaqoob\inst{7,10}} 

\offprints{J.S. Kaastra}

\institute{ SRON National Institute for Space Research, 
Sorbonnelaan 2, 3584 CA Utrecht, The Netherlands 
\and Department of Physics and Astronomy, 
  Georgia State University, Astronomy Offices, 
  One Park Place South SE, Suite 700, Atlanta, GA 30303 
\and Catholic University of America 
\and Laboratory for Astronomy and Solar Physics, 
NASA's Goddard Space Flight Center, Code 681, Greenbelt, MD 20771 
\and CASA, University of Colorado, 389 UCB, Boulder, CO 80309-0389, USA 
\and Joint Center for Astrophysics, University of Maryland, 
Baltimore County, 1000 Hilltop Circle, Baltimore, MD 21250 
\and Laboratory for High Energy Astrophysics, Code 660, 
 NASA's Goddard Space Flight Center, Greenbelt, MD 20771 
\and Physics Department, Lawrence Livermore National Laboratory, 
PO Box 808, L-41, Livermore, CA 94550 
\and Columbia Astrophysics Laboratory, Columbia University, 
538W. 120th Street, New York, NY 10027 
\and Department of Physics and Astronomy, Johns Hopkins University, 
 Baltimore, MD 21218 } 

\date{Received  / Accepted  } 

\abstract{
The Seyfert~1 galaxy NGC~5548 was observed for a week by Chandra using both the
HETGS and LETGS spectro\-meters. In this paper we study the time variability of
the continuum radiation. During our observation, the source showed a gradual
increase in flux over four days, followed by a rapid decrease and flattening of
the light curve afterwards. Superimposed upon these relatively slow variations
several short duration bursts or quasi-periodic oscillations occured with a
typical duration of several hours and separation between 0.6--0.9 days. The
bursts show a delay of the hard X-rays with respect to the soft X-rays of a few
hours. We interprete these bursts as due to a rotating, fluctuating hot spot at
approximately 10 gravitational radii; the time delay of the hard X-rays from
the bursts agree with the canonical picture of Inverse Compton scattering of
the soft accretion disk photons on a hot medium that is relatively close to the
central black hole.
}

\maketitle

\section{Introduction}

Active Galactic Nuclei (AGN) are well known for their violent environment. Gas
is being swallowed by the black hole, fed by a continuous supply of fresh
material through an accretion disk.  This process becomes visible as intense
high-energy radiation from the disk and its immediate surroundings, in
particular close to the black hole.  This radiation field may drive outflows
from the nucleus.  The detection of these outflows gives us a way to probe into
the inner nuclear regions.

Recent high-resolution X-ray spectra, starting with the first observation of
the Seyfert~1 galaxy \object{NGC~5548} (Kaastra et al. \cite{kaastra00}) and
high-resolution UV spectra (Crenshaw \& Kraemer \cite{crenshaw99}; Kriss et al.
\cite{kriss00}), allowed for an unprecedented study of the ionization and
dynamical structure of the outflowing photoionized winds. In the framework of a
study of these winds, we proposed to re-observe NGC~5548 in order to obtain
simultaneous high-resolution, high signal-to-noise X-ray and UV spectra.

NGC~5548 was observed for a full week with Chandra in January 2002 using both
the High Energy Transmission Grating Spectrometer (HETGS) and Low Energy
Transmission Grating Spectrometer (LETGS), with simultaneous UV observations
taken by the Space Telescope Imaging Spectrograph (STIS). The analysis of the
warm absorber is presented elsewhere (UV spectra: paper I, Crenshaw et al.
\cite{crenshaw03}; X-ray spectra: paper II, Steenbrugge et al.
\cite{steenbrugge04}). Limits to the spatial extent of the X-ray source were
discussed by Kaastra et al. (\cite{kaastra03}). In this paper we discuss the
continuum time variability of the source during this observation.

\section{Observations and data extraction}

The present observation of NGC~5548 was obtained in January 2002. The
observation was split over three orbits of the Chandra satellite. In the first
orbit (151~ks exposure time, start January 16, 2002) the HETG/ACIS-S
configuration was used, in the second (170~ks exposure, start January 18) and
third (171~ks exposure, start January 21) the LETG/HRC-S configuration was
used. The HETGS data were reduced using the standard CIAO software version
2.0b. The LETGS data were reduced using dedicated software as described in
Kaastra et al.(\cite{kaastra02}). The combined observation spans a time
interval of a full week. All times reported in this paper are given in seconds
relative to MJD~52290 (Jan 16, 2002, time 00:00:00).

\section{Time variability} 

\subsection{Light curve\label{sect:lc}} 

\begin{figure}[!ht]
\resizebox{\hsize}{!}{\includegraphics[angle=-90]{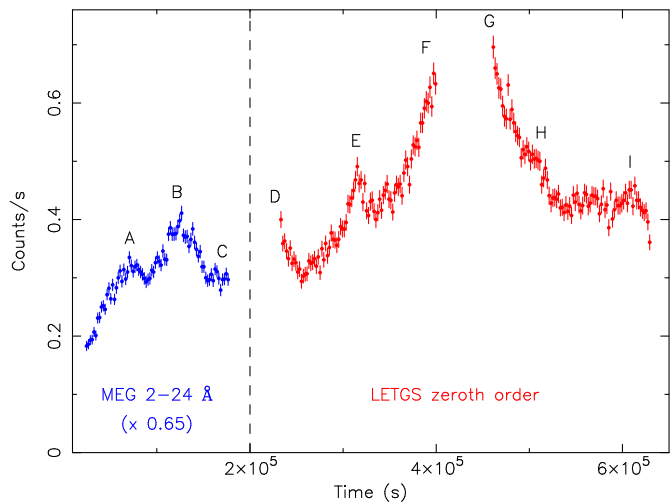}} 

\caption{X-ray light curve of NGC~5548.  The data have been binned in  bins of
2000~s. Times in this and all subsequent figures are measured relative to 
MJD~52290 (Jan 16, 2002, time 00:00:00).  The two data gaps are caused by  the
perigee passage of {\it Chandra}.  The left part of the plot contains the
2--24~\AA\ count rate as measured with the MEG in $+1$ and $-1$ order combined,
multiplied by a factor of 0.65 in order to be at approximately the same scale
as the zeroth order count rate with the LETGS (middle and right part).  We
estimate the uncertainty in this scale factor as $\sim$10~\%. Capitals indicate
the peaks discussed in the text. 
} 
\label{fig:zero} 
\end{figure} 

We constructed the LETGS zeroth order light curve of NGC~5548 by extracting all
events within a radius of 100 pixels from the location of the source. The
background, estimated from a nearby region, is negligible as compared to the
flux of the source, and shows no significant time variations. Note that due to
the use of the HRC-S detector, pile-up in the zeroth order can be neglected.

Fig.~\ref{fig:zero} shows the light curve extracted from the zeroth order of
the LETGS, combined with the first order MEG count rate, scaled to
approximately the corresponding LETGS level. We estimate the uncertainty in
this scaling factor to be $\sim 10$~\%, taking into account the absolute
calibration uncertainties of both instruments (in particular the zeroth order
effective area of the LETGS). The light curve is characterized by a rise during
four days from 0.2 to 0.7~c/s, with superimposed short duration (few hours)
fluctuations.  The characteristic exponential rise time toward the peak at
$t=400$~ks is a few days followed by a decay with a similar time scale up to
$t=520$~ks. After that time, the light curve remains approximately constant for
a day till the end of the observation. 

The light curve also exhibits a few  peaks, which we call here "bursts", and
which are labeled with the letters A through I in Fig.\ref{fig:zero}.  The
clearest example is the burst which  reaches its maximum at $t=314.2\pm 1.1$~ks
and lasts for about 20~ks (labeled "E" in  Fig.~\ref{fig:zero}).  Its peak
amplitude is about $0.092\pm 0.010$~c/s above the underlying rising flux of
0.369~c/s, i.e. an increase of 25~\%  of this steadily rising flux level. 

We note that bursts A, B and C are not only present in the MEG data, but are
also visible in the HEG data as well as the zeroth-order light curve of the
HETGS. This zeroth-order light curve, due to the use of the ACIS detector, has
strong pile-up effects which makes it less useful for quantitative analysis.
Therefore we concentrate here on the MEG first order data. The combined light
curves also show evidence for two very prominent short-lived bursts at either
side of burst B.

While bursts A, B and E are clearly recognized in Fig.~\ref{fig:zero}, the 
others are less obvious. Here we present our arguments why we think these other
bursts are significant. It is evident that the light curve has a local maximum 
between burst C and D. Instead of two or more separate bursts the data would 
also be consistent with a single peak during this interval. However, the
spectral variability (see Sect.~\ref{sect:specvar}) indicates a similar 
behavior at point C as at the start of the  better observed bursts: a maximum
in the soft flux while the hard flux is still  rising (Fig.~\ref{fig:hys2}).
 Therefore we tentatively assume that there are two bursts, C and D bracketing
the perigee passage (the instrumental background is small so this is not an
artifact induced by the Chandra orbit). Also, D is similar to the start of E
(see Fig.~\ref{fig:lcflux} and Fig.~\ref{fig:hys3}). This last figure shows the
same type of "hysteresis" for burst D (sequence 1--6 in the lower left of the
figure) as for burst E (the next sequence 1--6).

A similar situation holds for bursts F and G, which also bracket a perigee 
passage and where in particular peak F shows again the characteristic spectral
behavior for the onset of all other bursts: a flattening of the soft X-ray flux
while the hard X-ray flux is still rising. 

Evidence for burst H is most clearly seen as the shoulder in the light curve 
(Fig.~\ref{fig:zero}); however, its amplitude is small. Also burst I at the 
end of the observation has a small amplitude; this burst is weak and does not
show much evidence for delays between soft and hard bands. 

In summary, there is evidence for at least 7 bursts (A--G), some of them with 
relatively large amplitude, during the gradual rise in the first 400~ks of the
observation; there may be two other bursts after the peak maximum (H and I) but
they are less clear and weaker. 

We estimated the arrival times of the burst peaks by fitting Gaussians 
superimposed on a linear rise (Table~\ref{tab:burstpar}). There is no evidence
for true periodicity in the occurrence time of the bursts. However, the 
interval between the bursts has a characteristic time scale between 
$\sim$55--80~ks, with a slight tendency to increase starting from burst A to
the  maximum (around burst F and G). The duration of the bursts (as measured 
by the FWHM from the Gaussian fits) is $52\pm 9$, $54\pm 5$ and $18\pm 3$~ks 
for A, B and E, respectively. 

\begin{table}[!ht]
\caption{Burst parameters. Here $t_{\mathrm{peak}}$ is the time 
of maximum and $t_{\mathrm{next}}$ the time till the next burst maximum.} 
\label{tab:burstpar} 
\centerline{ 
\begin{tabular}{|lcr|} 
\hline 
Burst & $t_{\mathrm{peak}}$ (ks) & $t_{\mathrm{next}}$ (ks) \\ 
\hline 
A &    65$\pm$2   & 58 \\ 
B & 123.0$\pm$1.3 & $>$47 \\ 
C & $>$170        & $<$61 \\ 
D & 231$\pm$7     & 83 \\ 
E & 314.2$\pm$1.1 & 80 \\ 
F & 394$\pm$5     & $<$66 \\ 
G & $<$460        & $>$49 \\ 
H & 509$\pm$3     & 105 \\ 
I & 614$\pm$4     & - \\ 
\hline 
\end{tabular} 
} 
\end{table} 

\subsection{Continuum spectral variability\label{sect:specvar}} 

The spectrum of NGC~5548 is analyzed in detail by Steenbrugge et al.
(\cite{steenbrugge04}). Here we summarize their results relevant for the
present paper. The continuum is well approximated by the sum of a power law
component and a modified blackbody spectrum, a similar model as used by Kaastra
\& Barr (\cite{kaastra89}) in their analysis of the EXOSAT data of NGC~5548. In
addition, there is a warm absorber that produces both line and continuum
absorption. It appeared that while the continuum of NGC~5548 is changing, there
is no evidence for a significant change in the warm absorber, neither between
the HETGS and LETGS observation, nor between the present LETGS observation and
the first LETGS observation of December 1999, with the exception of \ion{O}{v}.
This simplifies our analysis of the continuum variability significantly. The
presence of a few narrow or broadened emission lines (for example the
\ion{O}{vii} forbidden line) does not affect our present broad-band analysis.

The zeroth order light curve of the LETGS contains no spectral information. 
However, taking our best-fit spectral model for the LETGS spectrum we estimate
that 90~\% of the zeroth order counts originate from photons between 2--50~\AA,
with a median photon wavelength of 18~\AA. We can get variability information
from the  first order spectral data, but due to the enhanced background of the
HRC-S detector the signal is noisier.  Therefore we need to  take larger time
bins than the 2000~s bins that we used in Fig.~\ref{fig:zero}. We have divided
the spectral data of both the HETGS and  LETGS in intervals of 12~ks. Utilizing
the high spectral resolution of the gratings, we then constructed broad-band
light curves which were corrected for the effective area of the instruments and
for Galactic absorption ($N_{\mathrm H} = 1.65\times 10^{24}$~m$^{-2}$). Since
the MEG effective area drops rapidly at long wavelengths, we restricted the MEG
wavelength band to $\lambda < 24$~\AA. 

\begin{figure} 
\resizebox{\hsize}{!}{\includegraphics[angle=-90]{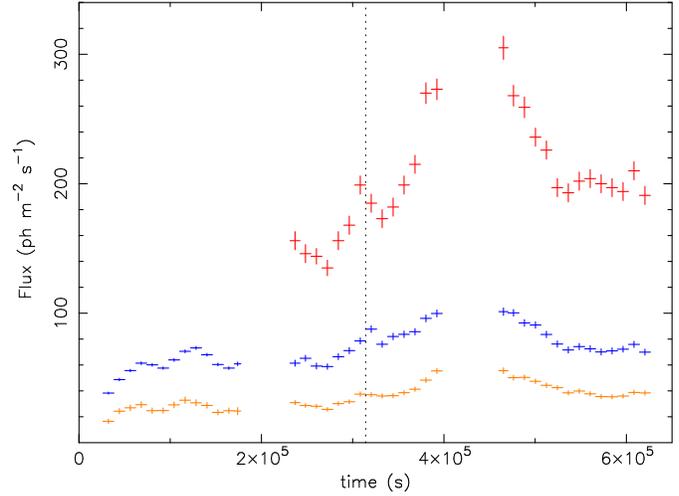}} 
\caption{X-ray light curve of NGC~5548 binned in 12000~s 
bins.  The fluxes have been corrected 
for Galactic absorption. From top to bottom: 18--50~\AA\ (soft band), 
2--12~\AA\ (hard band) and 18--24~\AA. Since the MEG is not sensitive to 
long wavelength photons, we have no data for the 18--50~\AA\ for the first 
part of the observation. The dotted vertical line indicates the maximum 
of the burst in the zeroth order light curve at $t=314.2$~ks. 
} 
\label{fig:lcflux} 
\end{figure} 

Fig.~\ref{fig:lcflux} shows the absorption-corrected light curves for three 
bands.  Our spectral modeling (Steenbrugge et al. \cite{steenbrugge04}) shows
that the hard band  (2--12~\AA) is entirely dominated by the power law
component, while the soft  band (18--50~\AA) is a mixture of power law and
modified blackbody emission. The figure shows that the variability is stronger
in the soft band:  see for example the sharp rise in soft flux shortly before
the maximum at $t=380$~ks. It is also evident that burst E at $t=314.2$
(indicated by the dotted line) is delayed in the hard band as compared to the
soft band, by about 1 bin (12~ks). 

\begin{figure} 
\resizebox{\hsize}{!}{\includegraphics[angle=-90]{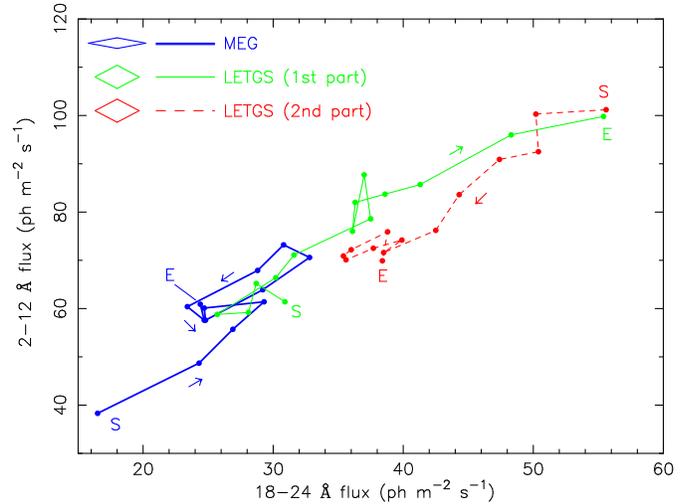}} 
\caption{Hard X-ray (2--12~\AA\ flux) versus soft X-ray (18--24~\AA\ flux) of 
NGC~5548. Fluxes have been corrected for Galactic absorption. Each data point
corresponds to a 12~ks interval. The start and end time bin of each part of the
observation are indicated by the "S" and "E" characters. Where needed, arrows
indicate the flow of the time sequence. The diamonds in the legend of the
figure indicate the error bars on individual data points (which are almost the
same for each data-point within each of the three parts of the observation). 
} 
\label{fig:hys2} 
\end{figure} 

In Fig.~\ref{fig:hys2} we show the relation between soft and hard flux, where 
for the soft flux we took the 18--24~\AA\ band in order to make use of the MEG 
data.  It is evident that during the HETGS observation and the first LETGS 
observation the flux in both bands is almost continuously rising, with at
least  four intermissions in the rise (at approximately the middle and end of
the HETGS  observation, and at the start and middle of the first LETGS
observation).  For each of these intermissions first the soft X-ray flux
starts decreasing, followed about 1 bin (12\,000~s) later by the hard X-ray
flux. 

\begin{figure} 
\resizebox{\hsize}{!}{\includegraphics[angle=-90]{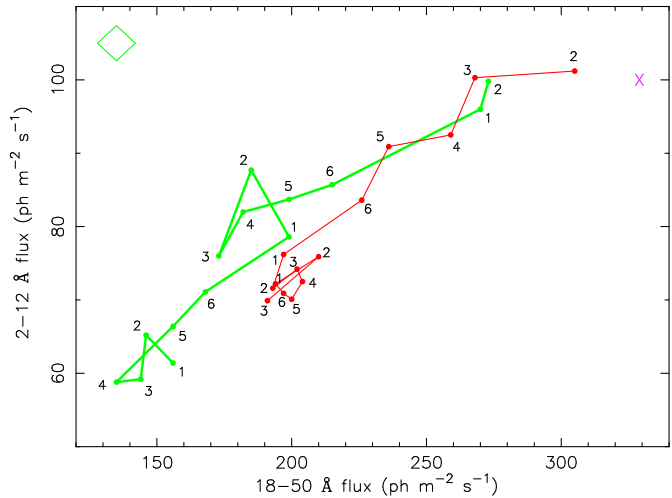}} 
\caption{Hard X-ray (2--12~\AA\ flux) versus soft X-ray (18--50~\AA\ flux) of 
NGC~5548 (LETGS observations only). Fluxes have been corrected for Galactic 
absorption. Each data point corresponds to a 12~ks interval. Data points 
are labeled following the sequence 1, 2, 3, 4, 5 and
6  repeatedly in time (6 times 12~ks is approximately the average time
interval  between burst peaks in the light curve). The diamond in the upper
left corner of the figure indicates the error bars on individual data points
(which are almost the same for each data-point in this figure). The "X" in the
upper right corner corresponds to the average LETGS spectrum of NGC~5548 during
the first observation with {\it Chandra} of this source in 1999 (Kaastra et al.
\cite{kaastra00}, \cite{kaastra02}). 
} 
\label{fig:hys3} 
\end{figure} 

This behaviour is even more evident from Fig.~\ref{fig:hys3}, where  only LETGS
data are plotted, so as to replace the 18--24~\AA\ flux by the 18--50~\AA\
flux. The flux in this latter band is on average 5.17$\pm$0.05 times larger
than in the more restricted band 18--24~\AA; the LETGS band shows an almost
strictly proportional relation between the flux in both soft bands. Due to the
limited statistics it is difficult to estimate the exact delay of the hard band
relative to the soft band; however by fitting locally Gaussians superimposed on
a linear rise for each band, using a higher time resolution version of our
light curves we find hard flux delays of $7\pm 3$, $5\pm3$, $3\pm 5$ and  $5\pm
3$~ks for bursts A, B, D and E, respectively. The weighted average is $5.4\pm
2.3$~ks. 

\begin{figure} 
\resizebox{\hsize}{!}{\includegraphics[angle=-90]{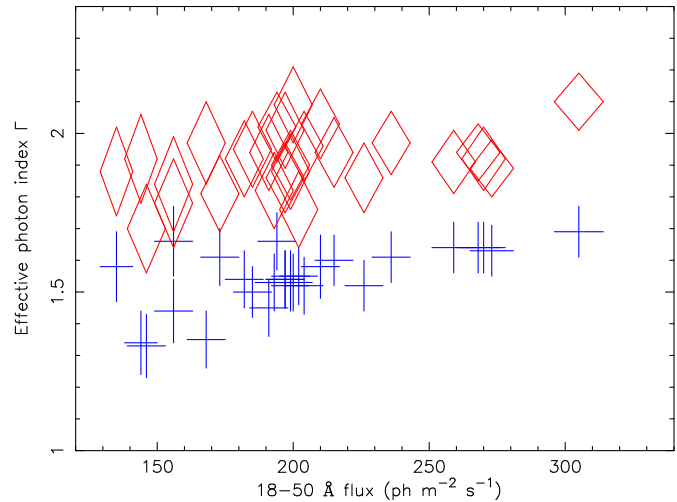}} 
\caption{Effective power law index $\Gamma$ in the 2--12~\AA\ band 
(crosses) and 18--50~\AA\ band (diamonds) versus 18--50~\AA\ Galactic 
absorption corrected flux for the LETGS data. 
} 
\label{fig:plvar} 
\end{figure} 

Further we note that our data show the well-known behaviour of the power law 
component: when the flux increases, the spectrum softens. We have fitted a 
simple power law to the fluxed, Galactic absorption corrected spectrum in the 
2--12~\AA\ band. This "effective" photon index ($\Gamma_{2-12}$) is not the
true photon index since that is affected by the warm absorber, but  given the
lack of spectral variability of the warm absorber (see Steenbrugge et al.
\cite{steenbrugge04}), it correlates well with the true photon index. This true
photon index of the power law component as deduced from the full spectral fit
to the total LETGS spectrum is 1.88 (Steenbrugge et al. \cite{steenbrugge04});
the time-averaged value of $\Gamma_{2-12}$ is 1.55. Given the lack of
variability of the warm absorber, one may add 0.33 to $\Gamma_{2-12}$ in
Fig.~\ref{fig:plvar} to get the true photon index. Since in the soft band the
continuum has a different nature (modified blackbody plus contributions from
the hard X-ray power law) the 18--50~\AA\ effective photon index
$\Gamma_{18-50}$ in Fig.~ref{fig:plvar} cannot be transformed easily into a
meaningfull physical photon index; it should be merely considered as a
convenient parameterisation of the effective spectral slope over this band.

Taking  $F_{18-50}$ the observed, Galactic absorption corrected flux in the
18--50~\AA\ band (in units of photons\,m$^{-2}$\,s$^{-1}$), we find a
correlation between  $\Gamma_{2-12}$ and $F_{18-50}$ (Fig.~\ref{fig:plvar}).
The best fit  parameters for a functional relation $\Gamma_{2-12} = a + b
F_{18-50}$ are $a=1.24\pm 0.12$ and $b=0.0015\pm 0.0006$. The relatively low
significance of the correlation is mainly due to the size of the errors on the
photon index, caused by the relatively small effective area of LETGS at the
highest energies. Contrary, the effective photon index  $\Gamma_{18-50}$
determined the same way but for the 18--50~\AA\ band does not show a strong
correlation with the soft X-ray flux (Fig.~\ref{fig:plvar}). In this case the
best fit parameters for $\Gamma_{18-50} = a + b F_{18-50}$ are $a=1.74\pm 0.15$
and $b=0.0009\pm 0.0007$. As is well known for NGC~5548 and other Seyfert
galaxies, our data do show the well known correlation between photon index
$\Gamma_{2-12}$ and flux $F_{2-12}$ for the power law component alone: the
photon index increases from 1.33 at $F_{2-12}=50$ to 1.66 for
$F_{2-12}=100$~photons\,m$^{-2}$\,s$^{-1}$.

Thus, the spectral shape in the soft band is less variable than in the  hard
band, despite the large flux variations in the soft band. 

\begin{figure} 
\resizebox{\hsize}{!}{\includegraphics[angle=-90]{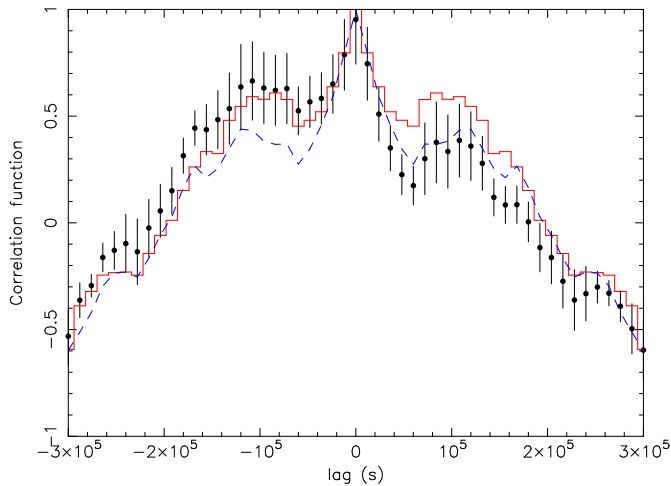}} 
\caption{Discrete correlation function (data points with error bars)
of the hard (2--12~\AA) with respect to the soft (18--24~\AA) band.
Positive lags correspond to hard X-rays following soft X-rays. The solid
line indicates the discrete autocorrelation function of the soft band,
the dashed line the discrete autocorrelation function of the hard band.
} 
\label{fig:dcf} 
\end{figure} 

Finally, we investigated the (cross)correlation of our data using the discrete
correlation function technique (Edelson \& Krolik \cite{edelson88}). In
Fig.~\ref{fig:dcf} we show our results. The figure shows a narrow peak with a
half-width of about 20--40~ks centered at zero lag. This corresponds to the
bursts in the light curve. Despite the fact that for individual bursts the hard
X-rays are delayed with respect to the soft X-rays (as can be seen in
Fig.~\ref{fig:hys2}), Fig.~\ref{fig:dcf} shows that on longer time scales the
hard X-rays tend to lead the soft X-rays. This can be seen from the stronger
correlation at a lag of $-100$~ks as compared to the correlation at a lag of
$+100$~ks. Note that due to the bin size of 12\,000~s used in
Fig.~\ref{fig:dcf} delays of individual bursts are invisible in that plot; but 
also in a higher time resolution correlation plot they remain invisible due to
the addition of noise from the non-burst parts of the lightcurve. Only our
hysteresis curves (Fig.~\ref{fig:hys2}--\ref{fig:hys3}) are able to demonstrate
the presence of these delays.

\section{Discussion} 

The lightcurve of NGC~5548 shows several bursts (Fig.~\ref{fig:zero}  and
Table~\ref{tab:burstpar}) with typical intervals between 55--80~ks. There is a
slight tendency of longer intervals around the maximum near burst F and G. The
spectra of these bursts show an average delay of the hard X-rays with respect
to the soft X-rays of 5.4$\pm$2.3~ks. 

Such a delay has been seen before in NGC~5548. Kaastra \& Barr 
(\cite{kaastra89}) found a delay of 4.6$\pm$1.2~ks in their correlation
analysis of the EXOSAT LE versus ME count rates from two long observations in
1986. Chiang et al. (\cite{chiang00}) find a delay of 13$\pm$6~ks in their 
correlation analysis of the  0.5--1~keV ASCA versus the $\sim$0.14--0.18~keV
EUVE Deep Survey (DS) count rates, for their coordinated  observations in 1998.
The delay of the 2--20~keV RXTE PCA  versus  EUVE count rate was larger,
34$\pm$11~ks, but there is less overlap between both light  curves as compared
to the ASCA and EUVE data sets. 

Interestingly, Dietrich et al. (\cite{dietrich01}) report an optical flare in
NGC~5548 on June 22, 1998 with a rise time of 1.8~ks and decay time less than
3.6~ks. This flare was visible in the U, B, V, R and I bands and had an
ampliude of $\sim$20~\%. However, no flare is visible in the simultaneous ASCA
and RXTE light curves published by Chiang et al. (\cite{chiang00}, see their
Fig~1; the peak of the optical flare (JD~2\,450\,987.37) is 197.2~ks after
their reference epoch). Unfortunately the optical flare occurred during a data
gap of the simultaneous EUVE observation. It is therefore hard to say if this
optical flare has a similar physical origin as the bursts that we observe.

Finally, we note that Haba et al.(\cite{haba03}) found 40~\% fluctuations in 
the DS light curve of NGC~5548 in 1996, without simultaneous variations in
either the ASCA count rate or spectrum. They conclude that if there is any
delay of hard X-rays with respect to soft X-rays, the delay time must be larger
than 60~ks. These soft X-ray fluctuations occur at the beginning of their
observation during a peak in the X-ray flux, and the DS light curve is
remarkably similar to our light curve starting near the maximum (bursts F and
G). 

We propose the following model for the bursts. We assume that the inner parts
of the accretion disk are affected by a thermal instability. This gives rise to
the gradual flux increase from the start of our observation to the maximum
around 400~ks after the start of our observation. We assume that there is a 
rotating hot spot at a distance $R = r GM/c^2$ from the black hole. This spot
may fluctuate in intensity and extend during its orbiting of the black hole.
Since the maximum emissivity of an accretion disk occurs near $r=10$, we assume
that $r$ is of this order of magnitude. 

Numerical simulations of accretion disks (for example Armitage \&
Reynolds \cite{armitage03}) indicate that hot spots can survive for a 
few orbital periods. 

The rotating hot spot produces soft X-rays by the same processes as the
surrounding parts of the disk emit soft X-rays. The details of this process are
not important for our present discussion, but it is clear that some form of
reprocessing or Comptonization must occur in the emitting regions, as the shape
of the soft X-ray spectrum changes little while its flux is increasing
significantly. In our model we assume that the hard X-rays are produced in a
different region, by Inverse Compton scattering of the seed photons from the
disk and hot spot.

\begin{figure} 
\resizebox{\hsize}{!}{\includegraphics[angle=-90]{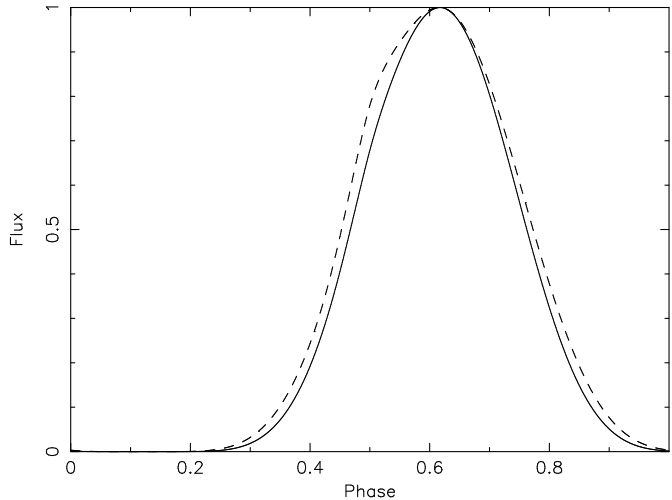}} 
\caption{Predicted pulse profile for a rotating spot around a Schwarzschild
black hole, at a distance of $R=10GM/c^2$ from the black hole. Solid
line: inclination 30$^{\circ}$; dashed line: inclination 50$^{\circ}$.
Phase 0.5 is defined as the instant when the spot is
behind the black hole. The FWHM of both profiles is 0.30 and 0.32, respectively.
} 
\label{fig:spot5548} 
\end{figure}

Analytical models for a rotating spot around a black hole were produced by
Karas (\cite{karas96}). At distances of $r\sim 10$, the light curve that would
be sinusoidal in the non-relativistic case is distorted by Doppler boosting and
general relativistic effects. The maximum is boosted while the minimum is
flattened in such a way that as a function of orbital phase the light curve is
almost  flat during half an orbit, and has a broad peak during the second half.
The maximum occurs at phase 0.6--0.7, when the hot spot is approaching the
observer (phase 0.5 is defined as the instant when the spot is behind the black
hole). The FWHM of the burst is about 30~\% of the orbital period. The above
holds for disk inclination angle $i<60^{\circ}$. For larger angles the light
curve is dominated by a narrower  peak at phase 0.5, caused by focusing of the
light by the black hole. However, here we assume that $i<60^{\circ}$. From the
presence of weak relativistic \ion{O}{viii} and \ion{N}{vii} line, Kaastra et
al. (\cite{kaastra02}) derived an inclination angle between 43--54$^{\circ}$,
in reasonable agreement with the  results of Yaqoob et al. (\cite{yaqoob01}) in
their re-analysis of ASCA data ($31^{\circ}\pm 8^{\circ}$). In
Fig.~\ref{fig:spot5548} we show an example of the predicted lightcurve for a
non-rotating black hole. The results for a Kerr black hole are very similar.

The observed interval between bursts of 55--80~ks then corresponds to the
orbital period of the hot spot. Irregularities are caused by fluctuations in
the emission from the other parts of the disk, as well as fluctuations in
brightness and extend of the spot. The expected FWHM of the bursts is
$\sim$30~\% of the orbital period, or between 15--25~ks. The observed duration
of the bursts is in agreement with this estimate, in particular for burst
E. A somewhat longer duration of bursts A and B near the onset of the
instability may be related to the initial formation of the spot over a larger
area of the disk.

Peterson \& Wandel (\cite{peterson00}) estimated the mass of the central black
hole in NGC~5548 to be  $(5.9 \pm 2.5) \times 10^{7}$M$_{\odot}$, based upon
reverberation mapping of the broad emission lines.  Using this mass range, an
orbital period of 55--80~ks corresponds to $r=$8--18 gravitational radii. In
our further estimates, we adopt $r=10$ and $M= 5.9 \times 10^{7}$M$_{\odot}$. 

Although the peak luminosity of the bursts reaches some 25~\% of the underlying
rising flux, it should be noted that the burst peaks are highly boosted by
Doppler and general relativistic effects. We estimate that the time-averaged
power of the bursts is only 5--10~\% of the total flux. This then implies a
typical size of a few gravitational radii for the hot spot.

The expected delay of the hard X-rays with respect to the soft X-rays depends 
upon the geometry of the emitting regions. There are two factors that may cause
a delay. If the Inverse Compton scattering of soft X-ray seed photons occurs
much further inward as compared to the location of the hot spot, a delay of
order $rGM/c^3$ or 3000~s may occur. The precise delay depends upon subtileties
like inclination angle, distance and geometry of the scattering region and
general relativistic effects. The details of such a model are beyond the scope
of this paper, however we note that the expected time delay is of the observed
order of magnitude. 

The second possible origin of a delay is retardation due to the random walk of
the soft X-ray seed photons in the scattering region, as they are being
up-scattered to hard X-rays. For typical photon indices and temperatures,
optical depths of the order of 2--3 are expected. This produces typical delays
of the order of 2--3 times the light travel time through the scattering region.
The probable size of the scattering region is less than $r$, the radius of the
soft X-ray emitting region, hence this may give an effect of the same order of
magnitude as the time delays due to the separation between hard and soft X-ray
emitting regions. Again, the details are very model-dependent. If this scenario
holds, then the hard X-ray source has an electron density of order
$10^{10}$~cm$^{-3}$.

Finally, we propose that around the maximum of the thermal instability, in the
middle of our LETGS observation, the hot spot may disappear, as we do not see
evidence for strong bursts after this maximum. From the hysteresis plots
(Figs.~\ref{fig:hys2} and \ref{fig:hys3}) we see that at the end of the
observation the continuum spectral properties differ significantly from the
spectral properties during the rise toward the peak, as is evident from the
different hard to soft X-ray flux ratio. A similar two state situation has been
found in \object{NGC~3783} by Netzer et al.(\cite{netzer03}). 

\section*{ACKNOWLEDGMENTS}

SRON is supported financially by NWO, the Netherlands Organization for
Scientific Research.


\begin{thebibliography}{}

\bibitem[2003]{armitage03}
Armitage, P.J., \& Reynolds, C.S., 2003, MNRAS, 341, 1041

\bibitem[2003]{chiang00}
Chiang, J., Reynolds, C. S., Blaes, O., M., et al., 2000, ApJ, 528, 292

\bibitem[1999]{crenshaw99}
Crenshaw, D.M., \& Kraemer, S.B., 1999, ApJ, 521, 572

\bibitem[2003]{crenshaw03}
Crenshaw, D.M., Kraemer, S.B., Gabel, J.R., et al., 2003, ApJ, 594, 116

\bibitem[2001]{dietrich01}
Dietrich, M., Bender, C.F., Bergmann, D.J., et al., 2001, A\&A, 371, 79

\bibitem[1988]{edelson88}
Edelson, R.A., \& Krolik, J.H., 1988, ApJ, 333, 646

\bibitem[2003]{haba03}
Haba, Y., Kunieda, H., Misaki, K., et al. 2003, ApJ, 599, 949

\bibitem[1989]{kaastra89}
Kaastra, J. S. \& Barr, P., 1989, A\&A, 226, 59

\bibitem[2000]{kaastra00}
Kaastra, J. S., Mewe, R., Liedahl, D. A., Komossa, S., 
Brinkman, A. C., 2000, A\&A, 354, L83

\bibitem[2002]{kaastra02}
Kaastra, J. S., Steenbrugge, K. C., Raassen, A. J. J., et al., 
2002a, A\&A, 386, 427

\bibitem[2003]{kaastra03}
Kaastra, J.S., Steenbrugge, K.C., Brinkman, A.C., et al., 2003,
in Active Galactic Nuclei: from central engine to host galaxy,
PASP Conf. Ser. 290, 101

\bibitem[1996]{karas96}
Karas, V., 1996, ApJ, 470, 743

\bibitem[2000]{kriss00}
Kriss, G.A., Green, R.F., Brotherton, M., et al., 2000, ApJ, 538, L17

\bibitem[2003]{netzer03}
Netzer, H., Kaspi, S., Behar, E., et al., 2003, ApJ, 599, 933

\bibitem[2000]{peterson00}
Peterson, B.M., \& Wandel, A.  2000, ApJ, 540, L13

\bibitem[2004]{steenbrugge04}
Steenbrugge, K.C., Kaastra, J.S., Crenshaw, D.M., et al., 2004, A\&A,
submitted

\bibitem[2001]{yaqoob01}
Yaqoob, T., George, I. M., Nandra, K., et al., 2001, ApJ, 546, 759

\end{thebibliography}
\end{document}